\documentclass{article}
\usepackage[T1]{fontenc} 
\usepackage[utf8]{inputenc} 
\usepackage{ismir,amsmath,cite,url}
\usepackage{graphicx}
\usepackage{color}

\usepackage{lipsum}
\usepackage{placeins}
\usepackage{float}
\usepackage{adjustbox}
\usepackage{amssymb}
\usepackage{multirow}
\usepackage{caption}
\title{Cover Detection Using Dominant Melody Embeddings}

\twoauthors
{Guillaume Doras} {\small Sacem \& Ircam Lab, CNRS, Sorbonne Université\\ \small 
     {\tt \small guillaume.doras@sacem.fr}}
 {Geoffroy Peeters} {\small LTCI, Telecom Paris, Institut Polytechnique de Paris\\ 
  {\tt \small geoffroy.peeters@telecom-paris.fr}}

\begin{document}

\maketitle
\begin{abstract}
Automatic cover detection -- the task of finding in an audio database all the covers of one or several query tracks -- has long been seen as a challenging theoretical problem in the MIR community and as an acute practical problem for authors and composers societies. Original algorithms proposed for this task have proven their accuracy on small datasets, but are unable to scale up to modern real-life audio corpora. On the other hand, faster approaches designed to process thousands of pairwise comparisons resulted in lower accuracy, making them unsuitable for practical use.

In this work, we propose a neural network architecture that is trained to represent each track as a single embedding vector. The computation burden is therefore left to the embedding extraction -- that can be conducted offline and stored, while the pairwise comparison task reduces to a simple Euclidean distance computation. We further propose to extract each track's embedding out of its dominant melody representation, obtained by another neural network trained for this task. We then show that this architecture improves state-of-the-art accuracy both on small and large datasets, and is able to scale to query databases of thousands of tracks in a few seconds.

\end{abstract}
\section{Introduction}\label{sec:introduction}
Covers are different interpretations of the same original musical work. They usually share a similar melodic line, but typically differ greatly in one or several other dimensions, such as their structure, tempo, key, instrumentation, genre, etc. 
Automatic cover detection -- the task of finding in an audio database all the covers of one or several query tracks -- has long been seen as a challenging theoretical problem in MIR. It is also now an acute practical problem for copyright owners facing continuous expansion of user-generated online content.

Cover detection is not \textit{stricto sensu} a classification problem: due to the ever growing amount of musical works (the classes) and the relatively small number of covers per work, the actual question is not so much ``to which work this track belongs to ?'' as ``to which other tracks this track is the most similar ?''.

Formally, cover detection therefore requires to establish a similarity relationship $S_{ij}$ between a query track $A_i$ and a reference track $B_j$. It implies the composite of a feature extraction function $f$ followed by a pairwise comparison function $g$, expressed as $S_{ij} = g(f(A_i), f(B_j))$. If $f$ and $g$ are independent, the feature extraction of the reference tracks $B_j$ can be done offline and stored. The online feature extraction cost is then linear in the number of queries, while pairwise comparisons cost without optimisation scales quadratically in the number of tracks \cite{humphrey2013data}. 

Efficient cover detection algorithms thus require a fast pairwise comparison function $g$. Comparing pairs of entire sequences, as DTW does, scales quadratically in the length of the sequences and becomes quickly prohibitive. At the opposite, reducing $g$ to a simple Euclidean distance computation between tracks embeddings is independent of the length of the sequences. In this case, the accuracy of the detection entirely relies on the ability of $f$ to extract the common musical facets between different covers.

In this work, we describe a neural network architecture mapping each track to a single embedding vector, and trained to minimize cover pairs Euclidean distance in the embeddings space, while maximizing it for non-cover pairs. We leverage on recent breakthroughs in dominant melody extraction, and show that the use of dominant melody embeddings yield promising performances both in term of accuracy and scalability. 

The rest of the paper is organized as follow: we review in \S\ref{sec:related_work} the main concepts used in this work. We detail our method in \S\ref{sec:method}, and describe and discuss in \S\ref{sec:preliminary_experiments} and \S\ref{sec:lookup_experiments} the different experiments conducted and their results. We finally present a comparison with existing methods in \S\ref{sec:comparison_soa}. We conclude with future improvements to bring to our method.

\section{Related work}\label{sec:related_work}

We review here the main concepts used in this study.

\subsection{Cover detection}\label{subsec:covers_detection}

Successful approaches in cover detection used an input representation preserving common musical facets between different versions, in particular dominant melody \cite{sailer2006finding, marolt2008mid, tsai2008using}, tonal progression -- typically a sequence of chromas  \cite{ellis2007identifyingcover, tsai2005query, gomez2006song, serra2009cross} or chords \cite{bello2007audio}, or a fusion of both \cite{foucard2010multimodal, salamon2013tonal}. Most of these approaches then computed a similarity score between pairs of melodic and/or harmonic sequences, typically a cross-correlation \cite{ellis2007identifyingcover}, a variant of the DTW algorithm \cite{tsai2005query, gomez2006song, serra2009cross, martin2012blast}, or a combination of both \cite{ravuri2010cover}. 

These approaches lead to good results when evaluated on small datasets -- at most a few hundreds of tracks, but are not scalable beyond due to their expensive comparison function. Faster methods have recently been proposed, based on efficient comparison of all possible subsequences pairs between chroma representations \cite{silva2016simple}, or similarity search between 2D-DFT sequences derived from CQTs overlapping windows \cite{seetharaman2017cover}, but remain too costly to be scalable to query large modern audio databases.

Another type of method has been proposed to alleviate the cost of the comparison function and to shift the burden to the audio features extraction function -- which can be done offline and stored. The general principle is to encode each audio track as a single scalar or vector -- its embedding -- and to reduce the similarity computation to a simple Euclidean distance between embeddings. Originally, embeddings were for instance computed as a single hash encoding a succession of pitch landmarks \cite{bertin2011large}, or as a vector obtained by PCA dimensionality reduction of a chromagram's 2D-DFT \cite{bertin2012large} or with locality-sensitive hashing of melodic excerpts \cite{marolt2008mid}. 

As for many other MIR applications, ad-hoc -- and somewhat arbitrary -- hand-crafted features extraction was progressively replaced with data-driven automatic feature learning \cite{humphrey2012moving}. Different attempts to learn common features between covers have since been proposed: in particular, training a $k$-means algorithm to learn to extract an embedding out of chromagram's 2D-DFT lead to significant results improvements on large datasets \cite{humphrey2013data}. Similar approaches, commonly referred to as \textit{metric learning} approaches, have been used in different MIR contexts, such as music recommendation \cite{mcfee2012learning, van2013deep}, live song identification \cite{tsai2016known}, music similarity search \cite{raffel2016pruning}, and recently cover detection \cite{qi2018triplet}.

\subsection{Metric learning}\label{subsec:metric_learning}

Although the concept can be traced back to earlier works \cite{baldi1993neural, bromley1994signature}, the term of metric learning was probably coined first in \cite{xing2003distance} to address this type of clustering tasks where the objective is merely to assess whether different samples are similar or dissimilar. It has since been extensively used in the image recognition field in particular \cite{he2016identity, song2016deep, sohn2016improved}.

The principle is to learn a mapping between the input space and a latent manifold where a simple distance measure (such as Euclidean distance) should approximate the neighborhood relationships in the input space. There is however a trivial solution to the problem, where the function ends up mapping all the examples to the same point. Contrastive Loss was introduced to circumvent this problem, aiming at simultaneously \textit{pulling} similar pairs together and \textit{pushing} dissimilar pairs apart \cite{hadsell2006dimensionality}.

However, when the amount of labels becomes larger, the number of dissimilar pairs becomes quickly intractable. It was moreover observed in practice that once the network has become reasonably good, negative pairs become relatively easy to discern, which stalls the training of the discriminative model. \textit{Pair mining} is the strategy of training the model only with hard pairs, i.e. positive (resp. negative) pairs with large (resp. small) distances \cite{simo2015discriminative}. Further improvement was introduced with the triplet loss, which is used to train a model to map each sample to an embedding that is closer to all of its positive counterparts than it is to all of its negative counterparts \cite{schroff2015facenet}. Formally, for all triplets \{$a$, $p$, $n$\} where $a$ is an anchor, and $p$ or $n$ is one of its positive or negative example, respectively, the loss to minimize is expressed as $\ell=\max(0, d_{\mathrm{ap}} + \alpha - d_{\mathrm{an}})$, where $\alpha$ is a margin and $d_{\mathrm{ap}}$ and $d_{\mathrm{an}}$ are the distances between each anchor $a$ and $p$ or $n$, respectively.

\subsection{Dominant melody extraction}\label{subsec:main_melody_extraction}

Dominant melody extraction has long been another challenging problem in the MIR community \cite{klapuri2006multiple, vincent2010adaptive, salamon2012melody}. A major breakthrough was brought recently with the introduction of a convolutional network that learns to extract the dominant melody out of the audio Harmonic CQT \cite{bittner2017deep}. The HCQT is an elegant and astute representation of the audio signal in 3 dimensions (time, frequency, harmonic), stacking along the third dimension several standard CQTs computed at different minimal multiple frequencies. Harmonic components of audio signal will thus be represented along the third dimension and be localized at the same location along the first and second dimensions. This representation is particularly suitable for melody detection, as it can be directly processed by convolutional networks, whose 3-D filters can be trained to localize in the time and frequency plan the harmonic components. 

In a recent work \cite{doras2019use}, we suggested in an analogy with image processing that dominant melody extraction can be seen as a type of image segmentation, where contours of the melody have to be isolated from the surrounding background. 
We have thus proposed for dominant melody estimation an adaptation of U-Net \cite{ronneberger2015u} -- a model originally designed for medical image segmentation -- which slightly improves over \cite{bittner2017deep}.

\begin{figure*}
\centerline{
\includegraphics[width=0.8\textwidth]{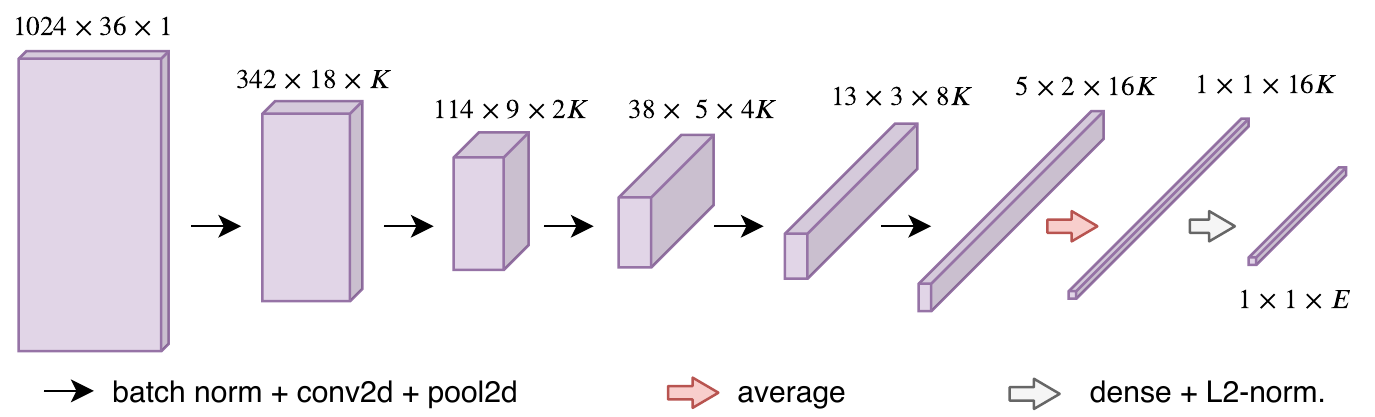}
}
 \caption{Convolutional model (time on the first dimension, frequency on the second dimension).}
 \label{fig:cnn_simple_2d}
\end{figure*}

\section{Proposed method}\label{sec:method}

We present here the input data used to train our network, the network architecture itself and its training loss.

\subsection{Input data}\label{subsec:method_input_data}

\begin{figure}[H]
\centerline{
\includegraphics[width=\columnwidth]{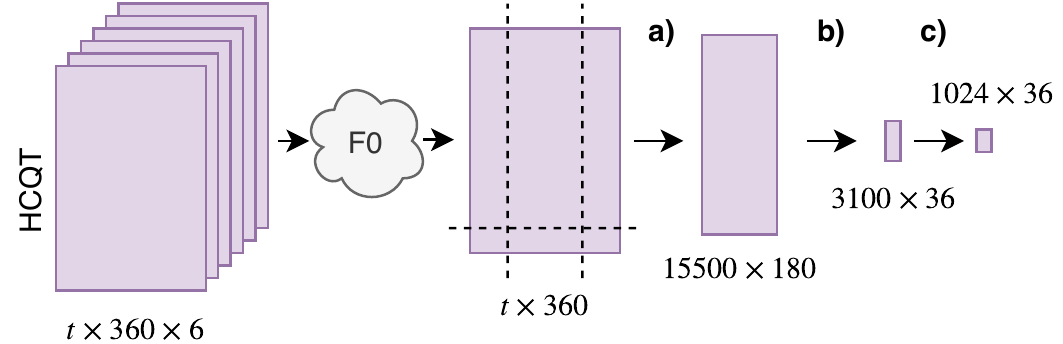}
}
 \caption{Input data pre-processing: dominant melody is extracted from HCQT, then a) F0 output is trimmed, b) downsampled by a factor 5 and c) resized time-wise to 1024 bins (time bins on first dimension, frequency bins on second dimension).}
 \label{fig:input_data}
\end{figure}

We have used as input data the dominant melody 2D representation (F0-CQT) obtained by the network we proposed in \cite{doras2019use}. The frequency and time resolutions required for melody extraction (60 bins per octave and 11 ms per time frame) are not needed for cover detection. Moreover, efficient triplet loss training requires large training batches, as we will see later, so we reduced data dimensionality as depicted on \figref{fig:input_data}.

The F0-CQT is \textbf{a)} trimmed to keep only 3 octaves around its mean pitch (180 bins along the frequency axis), and only the first 3 minutes of the track (15500 time frames) -- if shorter, the duration is not changed. The resulting matrix is then \textbf{b)} downsampled via bilinear 2D interpolation with a factor 5. On the frequency axis, the semi-tone resolution is thus reduced from five to one bin, which we considered adequate for cover detection. On the time axis, it is equivalent to a regular downsampling.

Finally, as the representation of different tracks with possibly different durations shall be batched together during training, the downsampled F0-CQT is \textbf{c)} shrunk or stretched along the time axis by another bilinear interpolation to a fixed amount of bins (1024). This operation is equivalent to a tempo change: for the 3 minutes trimmed, shrinking is equivalent to multiply the tempo by a factor 3. We argue here that accelerated or decelerated version of a cover is still a cover of the original track.

\FloatBarrier

\subsection{Model}\label{subsec:model}

The proposed model is a simple convolutional network pictured in \figref{fig:cnn_simple_2d}. As we are constrained by the input data shape, whose time dimension is much larger than its frequency dimension, only five layers blocks are needed. Each layer block consists of a batch normalization layer, a convolution layer with $3 \times 3$ kernels and a mean-pooling layer with a $3 \times 2$ kernel and $3 \times 2$ stride in order to reduce time dimensionality faster than frequency dimensionality. A dropout rate of $0.1, 0.1, 0.2$ and $0.3$ is applied to the blocks 2, 3, 4 and 5, respectively. 

The first convolutional layer has $K$ kernels, and this number is doubled at each level (i.e. the deeper layer outputs $2^4 K$-depth tensors). The penultimate layer averages along frequency and time axes to obtain a vector. A last dense layer outputs and L2-normalizes the final embedding vector of size $E$.

Our assumption behind the choice of this convolutional architecture is that we expect it to learn similar patterns in the dominant melody, at different scales (tempo invariance) and locations (key and structure invariance). 

\subsection{Objective loss}\label{subsec:triplet_loss}

We use a triplet loss with online semi-hard negative pairs mining as in \cite{schroff2015facenet}. In practice, triplet mining is done within each training batch: instead of using all possible triplets, each track in the batch is successively considered as the anchor, and compared with all its covers in the batch. For each of these positives pairs, if there are negatives such as $d_{\mathrm{an}} < d_{\mathrm{ap}}$, then only the one with the highest $d_{\mathrm{an}}$ is kept. If no such negative exist, then only the one with the lowest $d_{\mathrm{an}}$ is kept. Other negatives are not considered.

Model is fit with Adam optimizer \cite{kingma2014adam}, with initial learning rate at $1e^{-4}$, divided by 2 each time the loss on the evaluation set does not decrease after 5k training steps. Training is stopped after 100k steps, or if the learning rate falls below $1e^{-7}$. The triplet loss was computed using squared Euclidean distances (i.e. distances are within the $[0,4]$ range), and the margin was set to $\alpha=1$.

\subsection{Dataset}\label{subsec:dataset}

As metric learning typically requires large amount of data, we fetched from internet the audio of cover tracks provided by the SecondHandSongs website API\footnote{https://secondhandsongs.com/}. Only works with 5 to 15 covers, and only tracks lasting between 60 and 300 seconds where considered, for a total of $W=7460$ works and $T=62310$ tracks.

The HCQT was computed for those 62310 tracks as detailed in \cite{bittner2017deep}, i.e. with $f_{\mathrm{min}}$ = 32.7 Hz and 6 harmonics. Each CQT spans 6 octaves with a resolution of 5 bins per semi-tone, and a frame duration of \textasciitilde11 ms. The implementation was done with the Librosa library \cite{mcfee2015librosa}.

The dominant melody was extracted for these 62310 HCQT with the network we described in \cite{doras2019use}, and the output was trimmed, downsampled and resized as described in \S \ref{subsec:method_input_data}. 

\section{Preliminary experiments}\label{sec:preliminary_experiments}

We present here some experiments conducted to develop the system. The 7460 works were split into disjoint train and evaluation sets, with respectively 6216 and 1244 works and five covers per work. The evaluation set represents \textasciitilde20\% of the training set, which we considered fair enough given the total amount of covers. The same split has been used for all preliminary experiments.

\subsection{Metrics}\label{subsec:metrics}

Ideally, we expect the model to produce embeddings such that cover pair distances are low and non-cover pair distances are high, with a large gap between the two distributions. In the preliminary experiments, we have thus evaluated the separation of the cover pairs distance distribution $p_c(d)$ from the non-cover pairs distance distribution $p_{nc}(d)$ with two metrics:

- the ROC curve plots the true positives rate (covers, TPR) versus the false positive rate (non-covers, FPR) for different distance $d$ thresholds. We report the area under the ROC curve (AuC), which gives a good indication about the distributions separation. We also report the TPR corresponding to an FPR of 5\% (TPR@5\%), as it gives an operational indication about the model's discriminative power. 

- we also report the Bhattacharyya coefficient (BC), expressed as $\sum_d \sqrt{p_c(d)p_{nc}(d)}$, as it directly measures the separation between the distributions (smaller is better) \cite{bhattacharyya1943measure}.

\subsection{Influence of input data}\label{subsec:experiments_data_pre_processing}

\begin{figure}[H]
\centerline{
\includegraphics[width=\columnwidth]{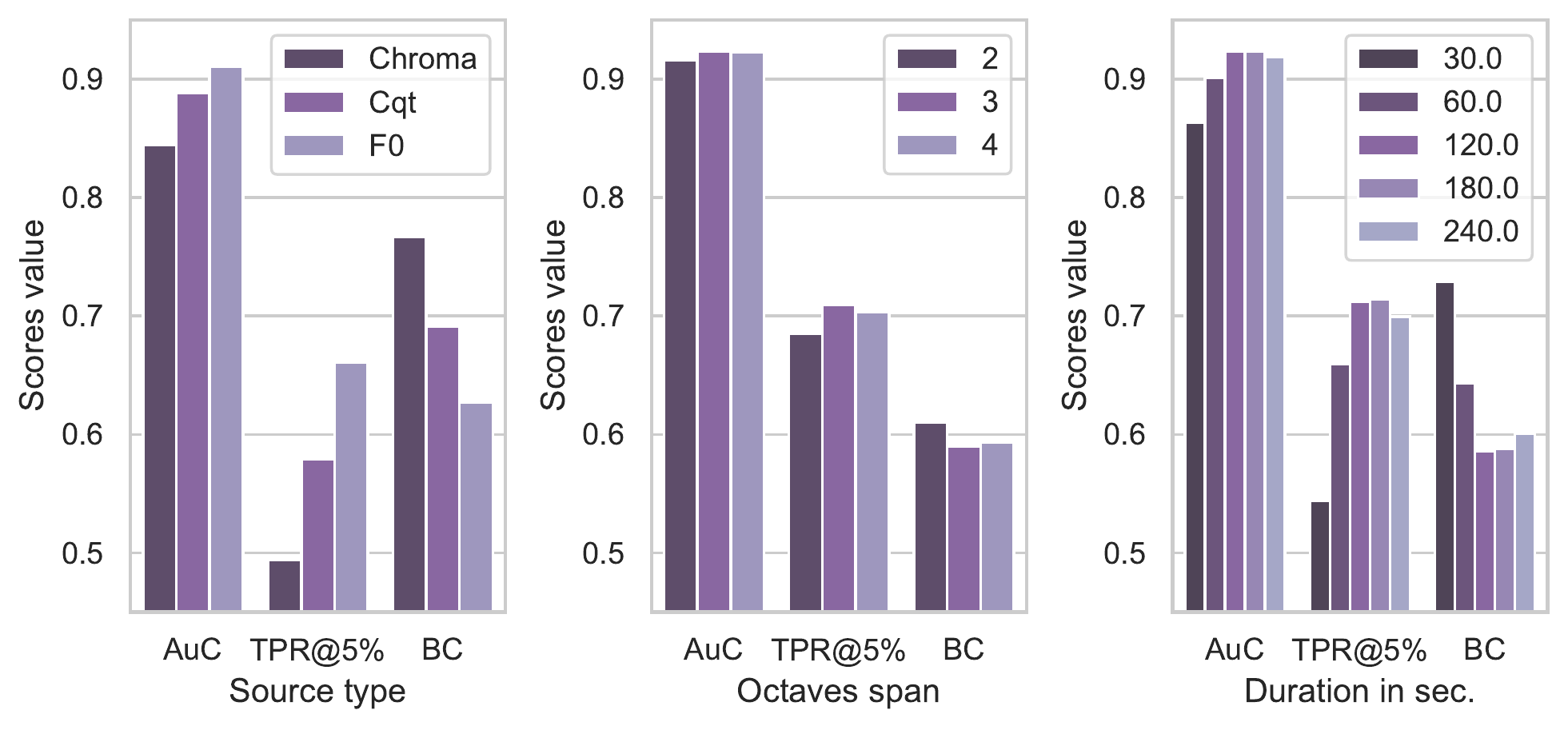}
}
 \caption{Scores obtained on evaluation set for a model trained - with chromas, CQT or F0 (left). - on the F0 with various octaves spans (middle). - on the F0 with various durations (right).}
 \label{fig:scores_for_sources_and_spans}
\end{figure}

We first compared the results obtained for different inputs data: chromas and CQT computed using Librosa \cite{mcfee2015librosa}, and the dominant melody computed as described in \ref{subsec:method_input_data}. As shown on \figref{fig:scores_for_sources_and_spans} (left), dominant melody yields the best results. It does not imply that melody features are more suited than tonal features for cover detection, but shows that convolutional kernels are better at learning similar patterns at different scales and locations across different tracks when the input data is sparse, which is not the case for chromas and CQT.

\FloatBarrier

Results obtained when trimming the F0-CQT with various octaves and time spans are also shown \figref{fig:scores_for_sources_and_spans}. It appears that keeping 3 octaves around the mean pitch of the dominant melody and a duration of 2 to 3 minutes yields the best results. Smaller spans do not include enough information, while larger spans generate confusion. 

All other results presented below are thus obtained with the dominant melody 2D representation as input data, and a span of 3 octaves and 180 seconds for each track.

\subsection{Influence of model and training parameters}\label{subsec:experiments_model_training_parameters}

\begin{figure}[H]
\centerline{
\includegraphics[width=\columnwidth]{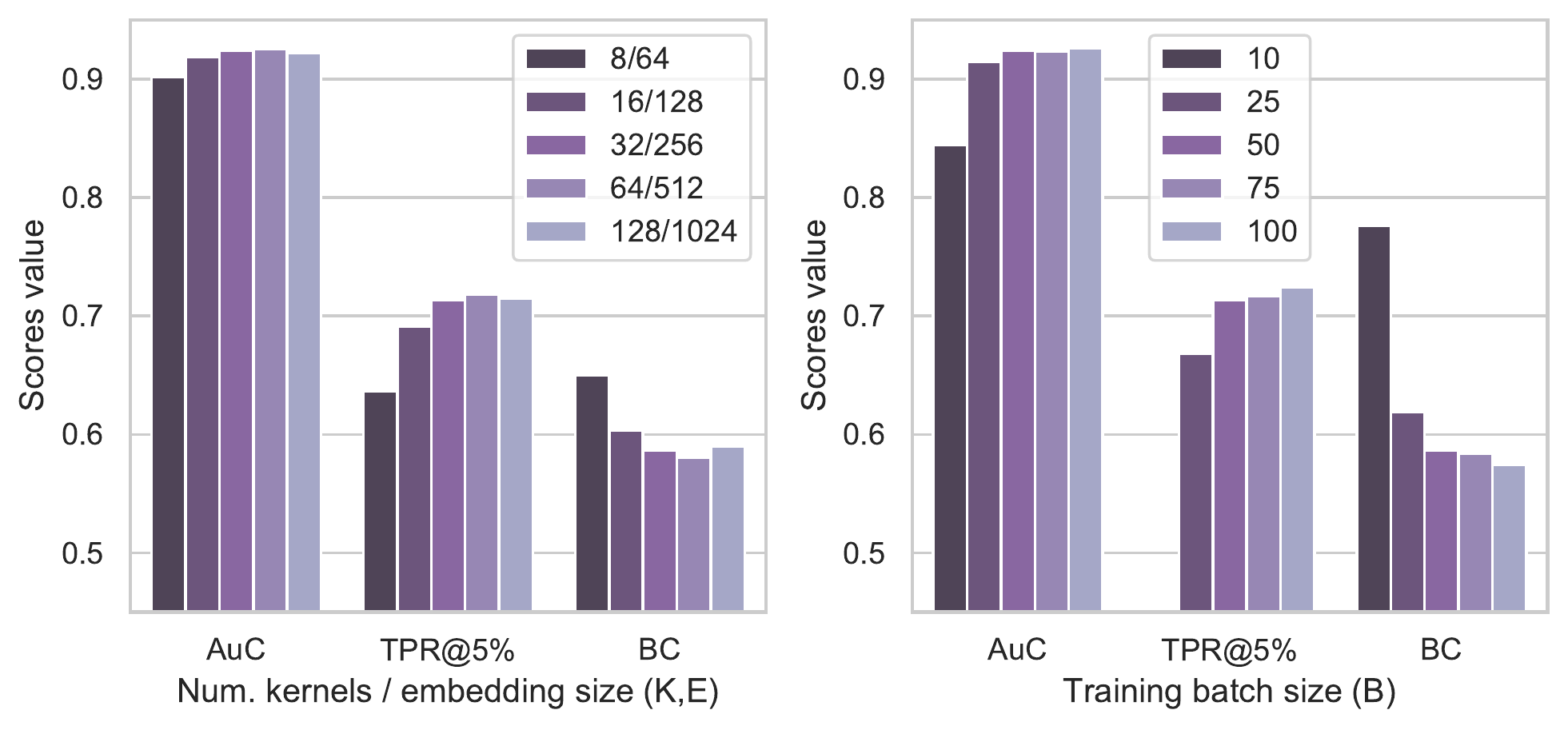}
}
 \caption{Scores obtained on evaluation set for a model trained - with various $K$/$E$ (left) - with various $B$ (right).}
 \label{fig:scores_for_batches_embeddings}
\end{figure}

We then compared the results obtained for different numbers of kernels in the first layer (K) and the corresponding sizes of the embeddings (E). As shown on \figref{fig:scores_for_batches_embeddings} (left), results improve for greater K, which was expected. However, increasing K above a certain point does not improve the results further, as the model has probably already enough freedom to encode common musical facets.

\FloatBarrier

We have then compared the results obtained for different sizes of training batches (B). As shown on \figref{fig:scores_for_batches_embeddings} (right), results improve with larger B: within larger batches, each track will be compared with a greater number of non-covers, improving the separation between clusters of works. A closer look at the distances shows indeed that the negative pairs distance distribution $p_{nc}(d)$ gets narrower for larger batches (not showed here). Due to GPU memory constraints, we have not investigated values above B=100.

All other results presented below are obtained with K=64, E=512 and B=100.

\section{Large scale lookup experiments}\label{sec:lookup_experiments}

We now present experiments investigating the realistic use case, i.e. large audio collections lookup. When querying an audio collection, each query track can be of three kinds: \textbf{a)} it is already present in the database, \textbf{b)} it is a cover of some other track(s) already in the database, or \textbf{c)} it is a track that has no cover in the database. The case \textbf{a)} corresponds to the trivial case, where the query will produce a distance equal to zero when compared with itself, while case \textbf{c)} corresponds to the hard case where neither the query or any cover of the query have been seen during training. We investigate here the case \textbf{b)}, where the query track itself has never been seen during training, but of which at least one cover has been seen during training.

\subsection{Metrics}

In these experiments, we are interested in measuring our method's ability to find covers in the reference set when queried with various unknown tracks. This is commonly addressed with the metrics proposed by MIREX\footnote{https://www.music-ir.org/mirex/wiki/2019} for the cover song identification task: the mean rank of first correct result (MR1), the mean number of true positives in the top ten positions (MT10) and the Mean Average Precision (MAP). We refer the reader to \cite{serra2007music} for a detailed review of these standard metrics. We also report here the TPR@5\%, already used in the premilinary experiments.

\subsection{Structuring the embeddings space}\label{subsec:large_experiments_structuring_space}

We study here the role of the training set in structuring the embeddings space, and in particular the role of the number of covers of each work. More precisely, we tried to show evidence of the \textit{pushing} effect (when a query is pushed away from all its non-covers clusters) and the \textit{pulling} effect (when a query is pulled towards its unique covers cluster).

\begin{figure}[H]
 \centerline{
 \includegraphics[width=0.73\columnwidth]{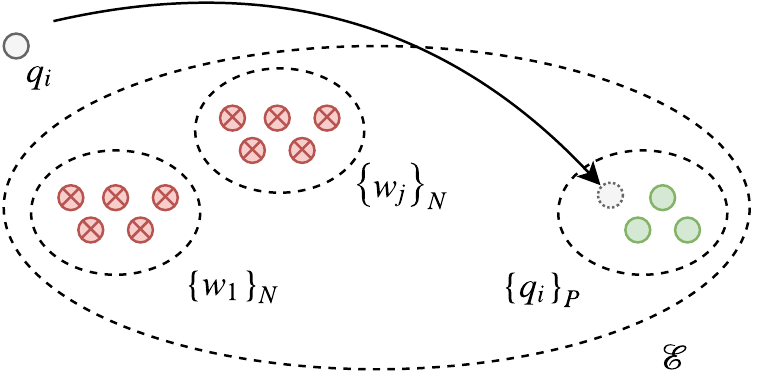}
 }
 \caption{Big dotted oval represents the embeddings space $\mathcal{E}$. Smaller dotted ovals represent work clusters of covers. Red crossed circles represent the positions in the manifold of the reference tracks $w_j$ that \textit{are not} covers of query track $q_i$ ($N$ per cluster). Green circles represent the positions of the tracks that \textit{are} covers of query track $q_i$ ($P$ per cluster).}
 \label{fig:train_test_scenario_1}
\end{figure}

To this aim, we built out of our dataset a query and a reference set. The query set includes $1244$ works with five covers each. The reference set includes $P$ of the remaining covers for each of the $1244$ query works, and $N$ covers for each other work not included in the query set ( \figref{fig:train_test_scenario_1}).

\begin{figure}[H]
\centerline{
\includegraphics[width=\columnwidth]{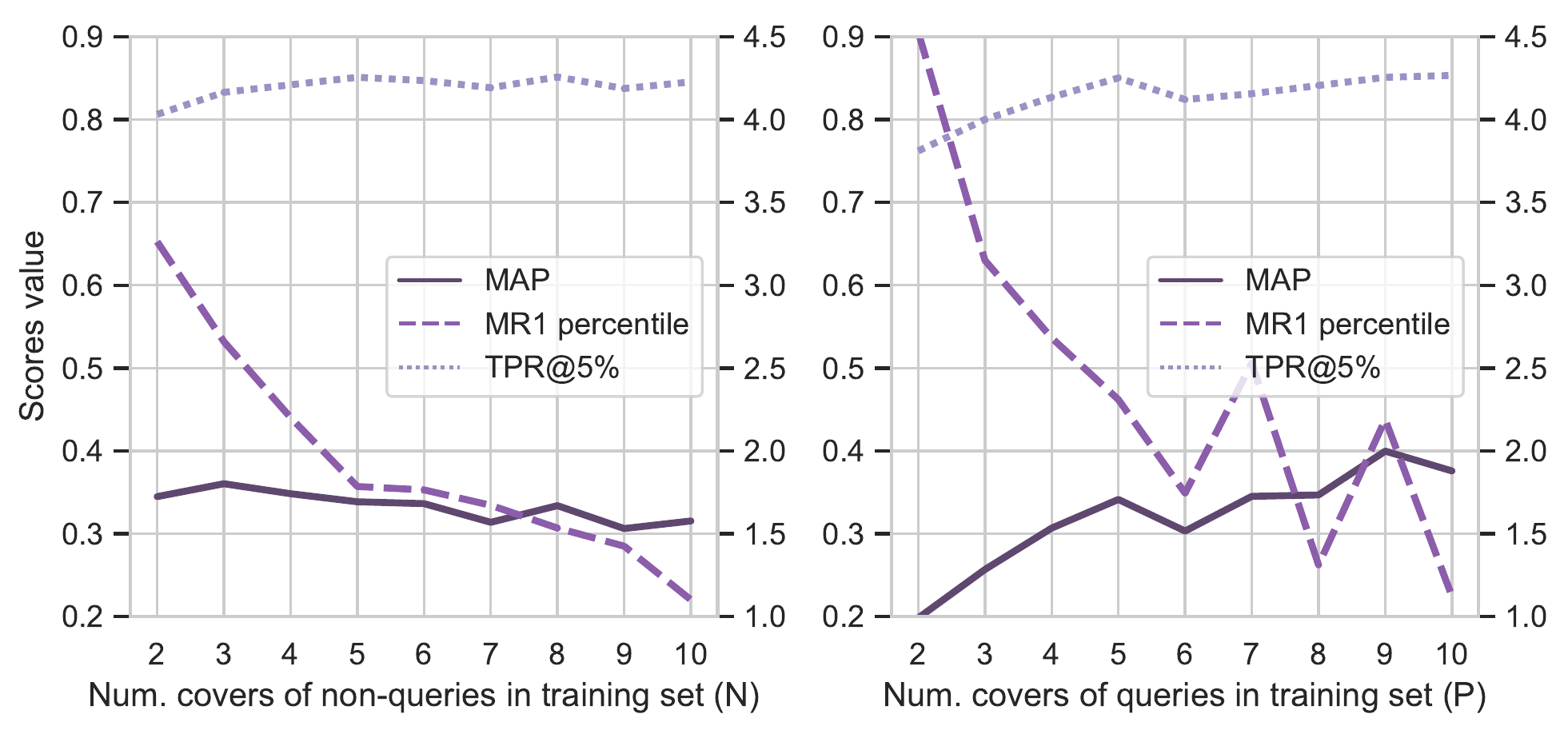}
}
 \caption{Query scores for different training/reference sets. Left: increasing N, number of covers of non-queries, P fixed. Right: increasing P, number of covers of queries, N fixed. (Y-axes scale MAP and TPR on the left, and the MR1 percentile on the right).}
 \label{fig:scores_for_training_covers_composition}
\end{figure}

\textbf{Pushing covers} We first train our model on the reference set with fixed $P$=5. We compute query tracks embeddings with the trained model, compute pairwise distances between query and reference embeddings, as well as the different metrics. We repeat this operation for different values of $N \in [2,...,10]$, and report results on \figref{fig:scores_for_training_covers_composition} (left). We report MR1's percentile (defined here as MR1 divided by the total of reference tracks, in percent) instead of MR1, because the number of reference tracks varies with $N$. 

\FloatBarrier

The MAP only slightly decreases as $N$ increases, which indicates that the precision remains stable, even though the number of examples to sort and to rank is increasing. Moreover, the MR1 percentile and the TPR@5\% clearly improve as $N$ increases. As $P$ is fixed, it means that the ranking and the separation between covers and non-covers clusters is improving as the non-queries clusters are consolidated, which illustrates the expected \textit{pushing} effect.

\textbf{Pulling covers} We reproduce the same protocol again, but now with $N$=5 fixed and for different values of $P \in [2,...,10]$. We report results on \figref{fig:scores_for_training_covers_composition} (right). It appears clearly that all metrics improve steadily as $P$ increases, even though the actual query itself has never been seen during training. As $N$ is fixed, this confirms the intuition that the model will get better in locating unseen tracks closer to their work's cluster if trained with higher number of covers of this work, which illustrates the expected \textit{pulling} effect.

\subsection{Operational meaning of $p_c(d)$ and $p_{nc}(d)$}\label{subsec:large_experiments_concrete_application}

We now investigate further the distance distributions of cover and non-cover pairs. To this aim, we randomly split our entire dataset into a query and a reference set with a 1:5 ratio (resp. 10385 and 51925 tracks). Query tracks are thus not seen during training, but might have zero or more covers in the reference set.\\ 

\begin{figure}[H]
 \centerline{
\includegraphics[width=\columnwidth]{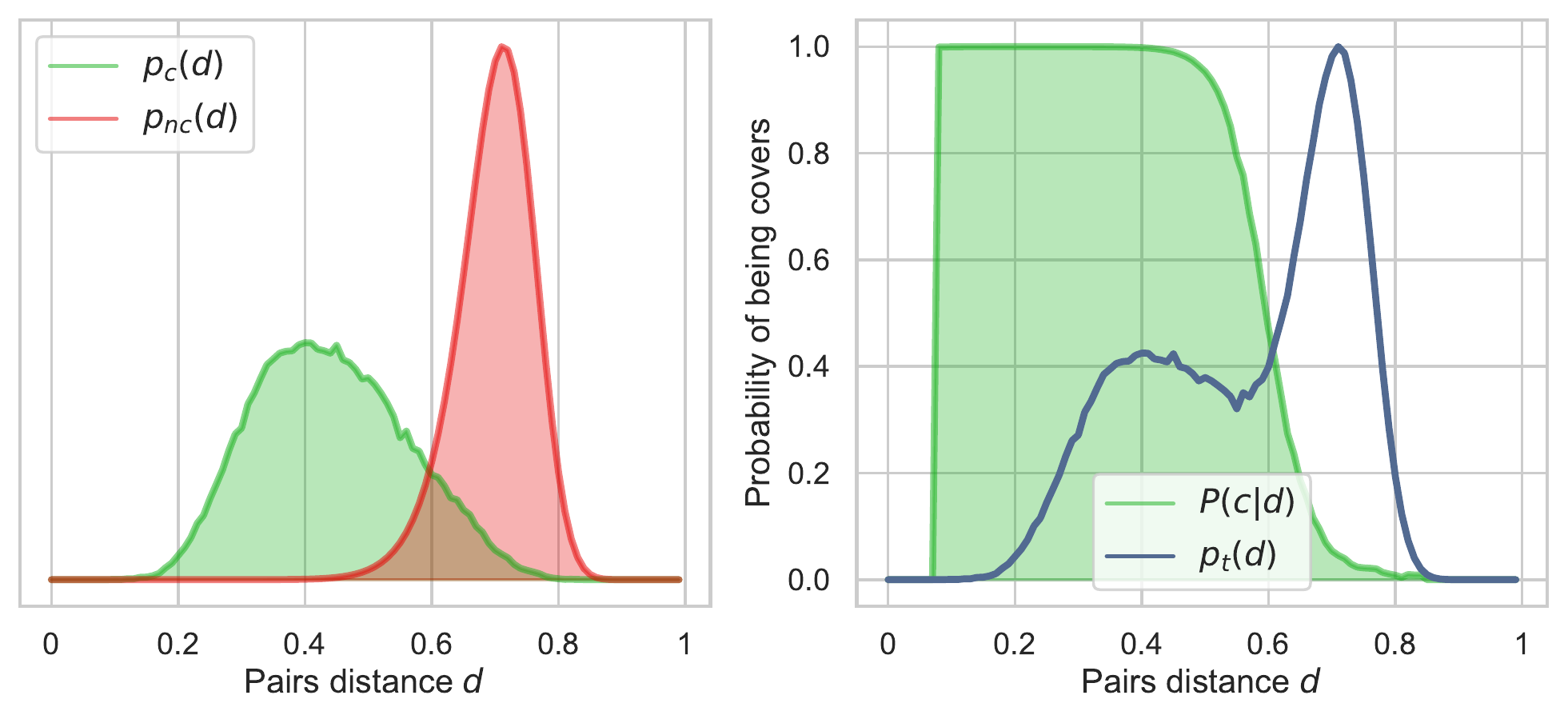}
 }
 \caption{Left: separation of $p_c(d)$ (green) and $p_{nc}(d)$ (red). Right: probability of being a covers pair given the distance $d$ (green) and total pairs distance distribution $p_c(d) + p_{nc}(d)$ (blue).}
 \label{fig:shs_distributions}
\end{figure}

\vspace{-0.5em}
\textbf{Covers probability} Computing queries vs. references pairwise distances gives the distributions $p_c(d)$ and $p_{nc}(d)$ shown on \figref{fig:shs_distributions} (left). Using Bayes' theorem, it is straightforward to derive from $p_c(d)$ and $p_{nc}(d)$ the probability for a pair of tracks to be covers given their distance $d$ (\figref{fig:shs_distributions}, right). This curve has an operational meaning, as it maps a pair's distance with a probability of being covers without having to rank it among the entire dataset.

\vspace{-0.5em}

\textbf{Easy and hard covers} We repeat the previous test five times with random splits, and report metrics in \tabref{tab:large_scale_experiments_dataset_results}. At first sight, MR1 and MT@10 could seem inconsistent, but a closer look at the results gives an explanation. To illustrate what happens, imagine a set of five queries where the first query ranks ten covers correctly in the first ten positions, e.g. because they are all very similar, while all other four queries have their first correct answer at rank 100. This would yield to MT@10=2.0, and MR1=80.2. This kind of discrepancy between MR1 and MT@10 reflects the fact that some works in our dataset have similar covers that are easily clustered, while other are much more difficult to discriminate. This can be observed on the positive pairs distribution $p_c(d)$ on \figref{fig:shs_distributions} (left), which is spread over a large range of distances.
\vspace{-0.2em}
\begin{table}[H]
\begin{adjustbox}{width=\columnwidth,center}
\begin{tabular}{l|c|c|c|c|c|c}
                            &MAP     &MT@10  &MR1 &BC   &AuC  &TPR@5\%\\
\hline
\multirow{2}{*}{Proposed}   &0.39   &2.90  &581 &0.40 &0.97 &0.88\\
                            &(<0.01) &(0.03)&(59)&(<0.01)&(<0.01)&(<0.01)\\
\end{tabular}
\end{adjustbox}
\caption{Results of query set lookup in reference set.}
\label{tab:large_scale_experiments_dataset_results}
\end{table}
\FloatBarrier

\section{Comparison with other methods}\label{sec:comparison_soa}

\subsection{Comparison on small dataset}\label{subsec:soa_small}
We first compared with two recent methods \cite{silva2016simple, seetharaman2017cover}, who reported results for a small dataset of 50 works with 7 covers each. The query set includes five covers of each work (250 tracks), while the reference set includes each work's remaining two covers (100 tracks). As this dataset is not publicly available anymore, we have mimicked it extracting randomly 350 tracks out of own dataset\footnote{\label{note1}We acknowledge that, strictly speaking, results can not be directly compared to rank methods, as original datasets of \cite{silva2016simple}, \cite{seetharaman2017cover} and \cite{humphrey2013data} are not available any longer. They however reflect the level of performance of the proposed method.}. 

Our data-driven model can however not be trained with only 100 tracks of the reference set, as it would overfit immediately. We have thus trained our model on our full dataset, with two different setups: \textbf{a)} excluding the 350 tracks reserved for the query and reference sets. \textbf{b)} excluding the 250 tracks of the query set, but including the 100 tracks of the reference set. We repeated this operation ten times for each setup, and report the mean and standard deviation on \tabref{tab:scores_vs_soa_small} for the same metrics used in \cite{silva2016simple, seetharaman2017cover}, as well as the p-value obtained by a statistical significance t-test carried out on results series. 

\begin{table}[H]
\begin{adjustbox}{width=\columnwidth,center}
\begin{tabular}{l|c|c|c}
                             &MAP               &P@10               &MR1\\
\hline
\cite{silva2016simple}       &0.591             &0.140              &7.910\\
\cite{seetharaman2017cover}  &0.648             &0.145              &8.270\\
Proposed \textbf{a)}         &\bf{0.675}        &\bf{0.165}         &\bf{3.439}\\
                             &(0.04), p=.29    &(0.005), p<.001    &(1.062), p<.001\\
Proposed \textbf{b)}         &\bf{0.782}        &\bf{0.179}         &\bf{2.618}\\
                             &(0.104), p<.01   &(0.014), p<.001    &(1.351), p<.001
                                    
\end{tabular}
\end{adjustbox}
\caption{Comparison between recent method \cite{silva2016simple, seetharaman2017cover} and our proposed method on a small dataset (precision at 10 P@10 is reported instead of MT@10. As there are only two covers per work in the reference set, P@10 maximum value is 0.2).}
\label{tab:scores_vs_soa_small}
\end{table}
\FloatBarrier

Our method significantly improve previous results: for the hardest case \textbf{a)} where the model has not seen any queries work during training, embeddings space has been sufficiently structured to discriminate the unseen works from the other training clusters (\textit{pushing} effect). For the easier case \textbf{b)}, the \textit{pulling} effect from the known queries covers provides further improvement.

\subsection{Comparison on large dataset}\label{subsec:soa_large}
We also compared with \cite{humphrey2013data}, who is to our knowledge the last attempt to report results for thousands of queries and references -- a more realistic use case.
This paper reported results on the SecondHandSong (SHS) subset of the MillionSong dataset (MSD) \cite{bertin2011million} for two experiments: \textbf{a)} only the training set of 12960 covers of 4128 works was used both as the query and reference sets. \textbf{b)} the SHS MSD test set of 5236 covers of 1726 works was used to query the entire MSD used as reference.

The SHS MSD is not available anymore. However, as our dataset has also been built from the SHS covers list, we consider that results can be compared\footnotemark[3]. We have therefore randomly generated out of our dataset a training and a test set mimicking the original ones. We trained our model on the training set, and perform the pairwise distances computation between the query and reference sets (as the query set is included in the reference set, we excluded for comparison the pairs of the same track). For experiment \textbf{b)}, we have used our entire dataset as reference set as we do not have one million songs. We have repeated this operation five times and report in \tabref{tab:scores_vs_soa_large} the mean and standard deviations for the same metrics used in \cite{humphrey2013data}, as well as MR1, MT@10 and the p-value of the t-test carried out.

\begin{table}[H]
\begin{adjustbox}{width=\columnwidth,center}
\begin{tabular}{c|l|c|c|c|c}
&        &MAP &MR &MT@10 &MR1 \\
\hline
\parbox[t]{2mm}{\multirow{3}{*}{\textbf{a)}}} 
&\cite{humphrey2013data}     &0.285             &1844 &- &- \\
&Proposed                    &\bf{0.936}        &\bf{78}        &2.010      &33   \\
&                            &(0.001), p<.001   &(6), p<.001    &(<0.001)    &(3)  \\
\hline
\parbox[t]{2mm}{\multirow{3}{*}{\textbf{b)}}} 
&\cite{humphrey2013data}   &0.134     &173117 &- &- \\

&Proposed                   &\bf{0.220}         &\bf{3865}      &1.622      &430 \\
&                           &(0.007), p<.001    &(81), p<.001   &(0.003)    &(19)  
\end{tabular}
\end{adjustbox}
\caption{Comparison between method \cite{humphrey2013data} and our proposed method on a large dataset (MR=Mean rank). For \textbf{b)}, the MR percentile should be compared, as our reference set does not have 1M tracks (6$^{th}$ vs. 17$^{th}$ percentile for \cite{humphrey2013data}).}
\label{tab:scores_vs_soa_large}
\end{table}
\FloatBarrier

Our method significantly improve previous results. For case \textbf{a)}, results are notably good, which is not surprising as the model has already seen all the queries during the training. Case \textbf{b)} is on the other hand the hardest possible configuration, where the model has not seen any covers of the queries works during training, and clusterisation of unseen tracks entirely relies on the \textit{pushing} effect.

As to our method's computation times, we observed on a single Nvidia GPU Titan XP for a \textasciitilde3 mn audio track: \textasciitilde \mbox{10 sec} for F0 extraction, \textasciitilde 1 sec for embeddings computation, and less than 0.2 sec for distances computation with the full dataset embeddings (previously computed offline).

\section{Conclusion}

In this work, we presented a method for cover detection, using a convolutional network which encodes each track as a single vector, and is trained to minimize cover pairs Euclidean distance in the embeddings space, while maximizing it for non-covers. We show that extracting embeddings out of the dominant melody 2D representation drastically yields better results compared to other spectral representations: the convolutional model learns to identify similar patterns in the dominant melody at different scales and locations (tempo, key and structure invariance).

We have also shown that our method scales to audio databases of thousands of tracks. Once trained for a given database, it can be used to assess the probability for an unseen track to be a cover of any known track without having to be compared to the entire database. We have finally shown that our method improves previous methods both on small and large datasets.

In the future, we plan to grow our training dataset to address the realistic use case where collections of millions of tracks should be queried: as for many other data-driven problems, will the cover detection problem be solved if the embeddings space is sufficiently structured?

\newpage
\textbf{Acknowledgements.}\\

The dominant melody representations of tracks used as training dataset in this work is available upon request.

We are grateful to Xavier Costaz at Sacem for his suggestions and support, and to Philippe Esling at Ircam for fruitful discussions.

\bibliography{./thesis.bib}

\begin{thebibliography}{10}

\bibitem{baldi1993neural}
Pierre Baldi and Yves Chauvin.
\newblock Neural networks for fingerprint recognition.
\newblock {\em Neural Computation}, 5(3):402--418, 1993.

\bibitem{bello2007audio}
Juan~Pablo Bello.
\newblock Audio-based cover song retrieval using approximate chord sequences:
  Testing shifts, gaps, swaps and beats.
\newblock In {\em Proceedings of ISMIR (International Society of Music
  Information Retrieval)}, 2007.

\bibitem{bertin2011large}
Thierry Bertin-Mahieux and Daniel~PW Ellis.
\newblock Large-scale cover song recognition using hashed chroma landmarks.
\newblock In {\em Proceedings of IEEE WASPAA (Workshop on Applications of
  Signal Processing to Audio and Acoustics)}, pages 117--120. IEEE, 2011.

\bibitem{bertin2012large}
Thierry Bertin-Mahieux and Daniel~PW Ellis.
\newblock Large-scale cover song recognition using the 2d fourier transform
  magnitude.
\newblock In {\em Proceedings of ISMIR (International Society of Music
  Information Retrieval)}, 2012.

\bibitem{bertin2011million}
Thierry Bertin-Mahieux, Daniel~PW Ellis, Brian Whitman, and Paul Lamere.
\newblock The million song dataset.
\newblock {\em Proceedings of ISMIR (International Society of Music Information
  Retrieval)}, 2011.

\bibitem{bhattacharyya1943measure}
Anil Bhattacharyya.
\newblock On a measure of divergence between two statistical populations
  defined by their probability distributions.
\newblock {\em Bull. Calcutta Math. Soc.}, 35:99--109, 1943.

\bibitem{bittner2017deep}
Rachel~M Bittner, Brian McFee, Justin Salamon, Peter Li, and Juan~P Bello.
\newblock Deep salience representations for f0 estimation in polyphonic music.
\newblock In {\em Proceedings of ISMIR (International Society of Music
  Information Retrieval)}, 2017.

\bibitem{bromley1994signature}
Jane Bromley, Isabelle Guyon, Yann LeCun, Eduard S{\"a}ckinger, and Roopak
  Shah.
\newblock Signature verification using a "siamese" time delay neural network.
\newblock In {\em Advances in Neural Information Processing Systems}, pages
  737--744, 1994.

\bibitem{doras2019use}
Guillaume Doras, Philippe Esling, and Geoffroy Peeters.
\newblock On the use of u-net for dominant melody estimation in polyphonic
  music.
\newblock In {\em International Workshop on Multilayer Music Representation and
  Processing (MMRP)}, pages 66--70. IEEE, 2019.

\bibitem{ellis2007identifyingcover}
Daniel~PW Ellis and Graham~E Poliner.
\newblock Identifyingcover songs with chroma features and dynamic programming
  beat tracking.
\newblock In {\em Proceedings of ICASSP (International Conference on Acoustics,
  Speech and Signal Processing)}. IEEE, 2007.

\bibitem{foucard2010multimodal}
R{\'e}mi Foucard, Jean-Louis Durrieu, Mathieu Lagrange, and G{\"a}el Richard.
\newblock Multimodal similarity between musical streams for cover version
  detection.
\newblock In {\em Proceedings of ICASSP (International Conference on Acoustics,
  Speech and Signal Processing)}. IEEE, 2010.

\bibitem{gomez2006song}
Emilia G{\'o}mez and Perfecto Herrera.
\newblock The song remains the same: identifying versions of the same piece
  using tonal descriptors.
\newblock In {\em Proceedings of ISMIR (International Society of Music
  Information Retrieval)}, 2006.

\bibitem{hadsell2006dimensionality}
Raia Hadsell, Sumit Chopra, and Yann LeCun.
\newblock Dimensionality reduction by learning an invariant mapping.
\newblock In {\em Proceedings of IEEE CVPR (Conference on Computer Vision and
  Pattern Recognition)}, volume~2, pages 1735--1742. IEEE, 2006.

\bibitem{he2016identity}
Kaiming He, Xiangyu Zhang, Shaoqing Ren, and Jian Sun.
\newblock Identity mappings in deep residual networks.
\newblock In {\em European conference on computer vision}, pages 630--645.
  Springer, 2016.

\bibitem{humphrey2012moving}
Eric~J Humphrey, Juan~Pablo Bello, and Yann LeCun.
\newblock Moving beyond feature design: Deep architectures and automatic
  feature learning in music informatics.
\newblock In {\em Proceedings of ISMIR (International Society of Music
  Information Retrieval)}, 2012.

\bibitem{humphrey2013data}
Eric~J Humphrey, Oriol Nieto, and Juan~Pablo Bello.
\newblock Data driven and discriminative projections for large-scale cover song
  identification.
\newblock In {\em Proceedings of ISMIR (International Society of Music
  Information Retrieval)}, 2013.

\bibitem{kingma2014adam}
Diederik~P Kingma and Jimmy Ba.
\newblock Adam: A method for stochastic optimization.
\newblock {\em arXiv preprint arXiv:1412.6980}, 2014.

\bibitem{klapuri2006multiple}
Anssi Klapuri.
\newblock Multiple fundamental frequency estimation by summing harmonic
  amplitudes.
\newblock In {\em Proceedings of ISMIR (International Society of Music
  Information Retrieval)}, 2006.

\bibitem{marolt2008mid}
Matija Marolt.
\newblock A mid-level representation for melody-based retrieval in audio
  collections.
\newblock {\em IEEE Transactions on Multimedia}, 10(8):1617--1625, 2008.

\bibitem{martin2012blast}
Benjamin Martin, Daniel~G Brown, Pierre Hanna, and Pascal Ferraro.
\newblock Blast for audio sequences alignment: a fast scalable cover
  identification.
\newblock In {\em Proceedings of ISMIR (International Society of Music
  Information Retrieval)}, 2012.

\bibitem{mcfee2012learning}
Brian McFee, Luke Barrington, and Gert Lanckriet.
\newblock Learning content similarity for music recommendation.
\newblock {\em IEEE Transactions on Audio, Speech, and Language Processing},
  20(8):2207--2218, 2012.

\bibitem{mcfee2015librosa}
Brian McFee, Colin Raffel, Dawen Liang, Daniel~PW Ellis, Matt McVicar, Eric
  Battenberg, and Oriol Nieto.
\newblock librosa: Audio and music signal analysis in python.
\newblock In {\em Proceedings of the 14th python in science conference}, pages
  18--25, 2015.

\bibitem{qi2018triplet}
Xiaoyu Qi, Deshun Yang, and Xiaoou Chen.
\newblock Triplet convolutional network for music version identification.
\newblock In {\em International Conference on Multimedia Modeling}, pages
  544--555. Springer, 2018.

\bibitem{raffel2016pruning}
Colin Raffel and Daniel~PW Ellis.
\newblock Pruning subsequence search with attention-based embedding.
\newblock In {\em Proceedings of ICASSP (International Conference on Acoustics,
  Speech and Signal Processing)}. IEEE, 2016.

\bibitem{ravuri2010cover}
Suman Ravuri and Daniel~PW Ellis.
\newblock Cover song detection: from high scores to general classification.
\newblock In {\em Proceedings of ICASSP (International Conference on Acoustics,
  Speech and Signal Processing)}. IEEE, 2010.

\bibitem{ronneberger2015u}
Olaf Ronneberger, Philipp Fischer, and Thomas Brox.
\newblock U-net: Convolutional networks for biomedical image segmentation.
\newblock In {\em International Conference on Medical image computing and
  computer-assisted intervention}, pages 234--241. Springer, 2015.

\bibitem{sailer2006finding}
Christian Sailer and Karin Dressler.
\newblock Finding cover songs by melodic similarity.
\newblock {\em MIREX extended abstract}, 2006.

\bibitem{salamon2012melody}
Justin Salamon and Emilia G{\'o}mez.
\newblock Melody extraction from polyphonic music signals using pitch contour
  characteristics.
\newblock {\em IEEE Transactions on Audio, Speech, and Language Processing},
  20(6):1759--1770, 2012.

\bibitem{salamon2013tonal}
Justin Salamon, Joan Serra, and Emilia G{\'o}mez.
\newblock Tonal representations for music retrieval: from version
  identification to query-by-humming.
\newblock {\em International Journal of Multimedia Information Retrieval},
  2(1):45--58, 2013.

\bibitem{schroff2015facenet}
Florian Schroff, Dmitry Kalenichenko, and James Philbin.
\newblock Facenet: A unified embedding for face recognition and clustering.
\newblock In {\em Proceedings of IEEE CVPR (Conference on Computer Vision and
  Pattern Recognition)}, pages 815--823, 2015.

\bibitem{seetharaman2017cover}
Prem Seetharaman and Zafar Rafii.
\newblock Cover song identification with 2d fourier transform sequences.
\newblock In {\em Proceedings of ICASSP (International Conference on Acoustics,
  Speech and Signal Processing)}. IEEE, 2017.

\bibitem{serra2007music}
Joan Serr{\`a}.
\newblock {\em Music similarity based on sequences of descriptors tonal
  features applied to audio cover song identification}.
\newblock PhD thesis, Universitat Pompeu Fabra, Spain, 2007.

\bibitem{serra2009cross}
Xavier Serra, Ralph~G Andrzejak, et~al.
\newblock Cross recurrence quantification for cover song identification.
\newblock {\em New Journal of Physics}, 11(9):093017, 2009.

\bibitem{silva2016simple}
Diego~F Silva, Chin-Chin~M Yeh, Gustavo Enrique de Almeida Prado~Alves Batista,
  Eamonn Keogh, et~al.
\newblock Simple: assessing music similarity using subsequences joins.
\newblock In {\em Proceedings of ISMIR (International Society of Music
  Information Retrieval)}, 2016.

\bibitem{simo2015discriminative}
Edgar Simo-Serra, Eduard Trulls, Luis Ferraz, Iasonas Kokkinos, Pascal Fua, and
  Francesc Moreno-Noguer.
\newblock Discriminative learning of deep convolutional feature point
  descriptors.
\newblock In {\em Proceedings of the IEEE International Conference on Computer
  Vision}, pages 118--126, 2015.

\bibitem{sohn2016improved}
Kihyuk Sohn.
\newblock Improved deep metric learning with multi-class n-pair loss objective.
\newblock In {\em Advances in Neural Information Processing Systems}, pages
  1857--1865, 2016.

\bibitem{song2016deep}
Hyun~Oh Song, Yu~Xiang, Stefanie Jegelka, and Silvio Savarese.
\newblock Deep metric learning via lifted structured feature embedding.
\newblock In {\em Proceedings of IEEE CVPR (Conference on Computer Vision and
  Pattern Recognition)}, pages 4004--4012. IEEE, 2016.

\bibitem{tsai2016known}
TJ~Tsai, Thomas Pr{\"a}tzlich, and Meinard M{\"u}ller.
\newblock Known artist live song id: A hashprint approach.
\newblock In {\em Proceedings of ISMIR (International Society of Music
  Information Retrieval)}, 2016.

\bibitem{tsai2005query}
Wei-Ho Tsai, Hung-Ming Yu, Hsin-Min Wang, et~al.
\newblock Query-by-example technique for retrieving cover versions of popular
  songs with similar melodies.
\newblock In {\em Proceedings of ISMIR (International Society of Music
  Information Retrieval)}, 2005.

\bibitem{tsai2008using}
Wei-Ho Tsai, Hung-Ming Yu, Hsin-Min Wang, and Jorng-Tzong Horng.
\newblock Using the similarity of main melodies to identify cover versions of
  popular songs for music document retrieval.
\newblock {\em Journal of Information Science \& Engineering}, 24(6), 2008.

\bibitem{van2013deep}
Aaron Van~den Oord, Sander Dieleman, and Benjamin Schrauwen.
\newblock Deep content-based music recommendation.
\newblock In {\em Advances in neural information processing systems}, pages
  2643--2651, 2013.

\bibitem{vincent2010adaptive}
Emmanuel Vincent, Nancy Bertin, and Roland Badeau.
\newblock Adaptive harmonic spectral decomposition for multiple pitch
  estimation.
\newblock {\em IEEE Transactions on Audio, Speech and Language Processing},
  18(3):528--537, 2010.

\bibitem{xing2003distance}
Eric~P Xing, Michael~I Jordan, Stuart~J Russell, and Andrew~Y Ng.
\newblock Distance metric learning with application to clustering with
  side-information.
\newblock In {\em Advances in Neural Information Processing Systems}, pages
  521--528, 2003.

\end{thebibliography}

\end{document}